\documentclass[aps,superscriptaddress,twocolumn,floats,tightenlines,balancelastpage,floatfix]{revtex4}
\input epsf
\usepackage{amsmath,amssymb}
\usepackage{graphicx}
\newcommand{\bo}{\raise-1mm\hbox{\Large$\Box$}}              % D'Alembertian
\begin{document}

\title{Lower Limit to the Scale of an Effective Theory of Gravitation}
\author{R.~R. Caldwell} 
\affiliation{Department of Physics \& Astronomy, Dartmouth College, 
Hanover, NH 03755}
\author{Daniel Grin} 
\affiliation{California Institute of Technology, Mail Code 130-33, Pasadena, CA
91125}
 
%\date{\today}

%%%%%%%%%%%%%%%%%%%%%%%%%%%%%%%%%%%%%%%%%%%%%%%%%%%%%%%%%%%%%%%%%%%%%%%%%%%%%%%%%%%%%%%%%%%%%%%%%%
\begin{abstract} 
An effective quantum theory of gravitation in which gravity weakens at energies higher than $\sim
10^{-3}~\rm{eV}$ is one way to accommodate the apparent smallness of the cosmological constant.
Such a theory predicts departures from the Newtonian inverse-square force law on distances
below $\sim 0.05$ mm. However, it is shown that this modification also leads to changes in the
long-range behavior of gravity and is inconsistent with observed gravitational lenses.
\end{abstract}
\maketitle
  
%%%%%%%%%%%%%%%%%%%%%%%%%%%%%%%%%%%%%%%%%%%%%%%%%%%%%%%%%%%%%%%%%%%%%%%%%%%%%%%%%%%%%%%%%%%%%%%%%%

The discovery of the cosmic acceleration \cite{SN:1998} has prompted speculations of new physics.
A leading hypothesis is the existence of a cosmological  constant, responsible for the
accelerated expansion. The milli-eV energy scale implied by this phenomenon is  difficult to
understand in terms of a fundamental theory \cite{Lambda}. The validity of
Einstein's general theory of relativity (GR) on cosmological scales has thus come under suspicion. A
novel solution to this problem might be achieved if GR is a low-energy effective theory in which
gravity weakens at some energy scale. In an effective theory of gravity there may exist a
threshold, $\mu$, beyond which gravitons cannot mediate momentum transfers. This behavior may be
due to a ``fat"  graviton, a minimal length scale associated with quantum gravity, or possibly
nonlinear effects which filter out high-frequency interactions 
\cite{Zee:2003dz,Sundrum:2003jq,Sundrum:2003tb,Hossenfelder:2006cw,paddy,reuter1,reuter2,Dvali:2007kt}.
Such theories offer a novel solution to the cosmological constant problem by regulating the
contribution of vacuum fluctuations to the cosmological constant.  However, we will show that
this mechanism may have already been explored and ruled out by gravitational lensing on
cosmological scales.

We estimate the energy scale of an effective theory of gravitation by matching the predicted
quantum vacuum energy density with the energy density of a cosmological constant, $\Lambda$, 
necessary to explain the accelerated cosmic expansion. Following Zeldovich
\cite{Zeldovich:1967gd}, the gravitating energy density of the particle physics vacuum
as due to $N$ equivalent, massless scalar particles, is
\begin{equation}
\rho_\Lambda =  \frac{N}{2} \int \frac{d^3 k}{(2 \pi \hbar)^3}\, k c \, f(k).
\end{equation}
We introduce the function $f(k) = {\rm e}^{-k/\mu}$ to regulate the momentum at the vertex where
vacuum bubbles connect to gravitons in order to limit the gravitating energy density. We refer
to $\mu$ as a ``cutoff" scale in the sense that the standard gravitational interactions are
severely weakened above this scale.  We match $\rho_\Lambda = \Omega_\Lambda \rho_{crit}$ and
obtain $\mu = 0.0048 (\Omega_\Lambda h^2
/N)^{1/4}$~eV/c as the desired cutoff scale. Current measurements give $\Omega_\Lambda h^2 = 0.34
\pm 0.04\,(1\sigma)$ (see Ref.~\cite{Roos:2005ra} and references therein) so that $\mu = 0.0037
(1 \pm 0.03)/N^{1/4}$~eV/c. We now examine the consequences of this cutoff.

We consider weak gravitational fields described by a linearized, effective quantum theory of
gravity \cite{QGravity}. The interaction Lagrangian at lowest order is
\begin{equation}
{\cal L}_I = -\frac{1}{2} \kappa h_{\mu\nu} T^{\mu\nu}
\label{Lint}
\end{equation}
where $\kappa = \sqrt{32 \pi G}$, $h_{\mu\nu}$ is the graviton field, and $T^{\mu\nu}$ is the
stress-energy tensor of the gravitating sources. Here, we introduce
an exponential cutoff at $\mu$ on graviton momenta. 

Short-distance gravitational phenomena below the length $\ell_0 = \hbar/\mu \sim 0.05$~mm are
affected by such a cutoff, which we impose on the graviton four-momentum $q^\mu$ so that
$q^{2}\equiv q^\mu q_\mu < \mu^2$. For real gravitons, $q^\mu q_\mu=0$ and so the constraint is
trivially satisfied. For virtual gravitons, the cutoff may be imposed
by suppressing the graviton propagator in the ultraviolet \cite{khouryfade}:
${1}/{q^{2}}\to  {\mathcal{G}\left(q^{2}/\mu^{2}\right)}/{q^{2}},$
where $\mathcal{G}$ is a function of the graviton momentum. For example, our exponential cutoff
follows if $\mathcal{G}\left(x\right)=e^{-\sqrt{x}}$. Such a modified propagator follows
naturally from modified gravitational Lagrangians. This is clear upon inspection of
the weak-field, Coulomb gauge, gravitational Lagrangian for a ``fading gravity" model \cite{khouryfade}:
\begin{equation}
\mathcal{L}_g=2\left(h^{\alpha \beta}-\frac{1}{2}\eta^{\alpha \beta}h\right)
\mathcal{G}^{-1}\left(\Box/\mu^{2}\right)\Box h_{\alpha \beta},
\label{fadgrav}
\end{equation}
where $\Box$ is the D'Alembertian operator. The sum of (\ref{Lint}) and (\ref{fadgrav}) can be
used to obtain the weak-field equations of motion.

An exponential cutoff to the momentum-space integral for the virtual gravitons
exchanged between two static masses, $m_1$ and $m_2$, changes the Newtonian potential to
\begin{eqnarray}
V &=& -8 \pi G m_1 m_2 \int \frac{d^3 q}{(2 \pi)^3 \hbar} 
\frac{1}{2 q^2}e^{\frac{i}{\hbar} \vec q \cdot (\vec x_1 - \vec x_2)} \times f(q) \cr
&=& -\frac{G m_1 m_2}{r} \times \frac{2}{\pi}\arctan{\frac{r}{\ell_0}}.
\label{vpot}
\end{eqnarray}
Relativistic corrections to the  potential are similarly modified
\cite{Corinaldesi1956,Barker1966}. The above expression asymptotes to the standard result for
$r\gg\ell_0$ but reaches a finite minimum as $r/\ell_0 \to 0$. Hence, static masses become free
of gravitation at short distances.

The possibility of new gravitational phenomena at submillimeter distances has motivated
laboratory tests of the Newtonian force law
\cite{Hoyle:2000cv,Chiaverini:2002cb,Long:2003dx,Hoyle:2004cw,Smullin:2005iv,Kapner:2006}. These
experiments look for departures from the Newtonian force law, which are interpreted as bounds on
a Yukawa-type modification of the potential,
$ V = -\frac{G m_1 m_2}{r}\times\left(1 + \alpha {\rm e}^{-r/\lambda}\right).$
The potential (\ref{vpot}) roughly corresponds to $\alpha \sim -1$ and $\lambda \sim \ell_0$.
Recent measurements show that the Newtonian force law holds down to $56\mu$m for $|\alpha|=1$ so
that $\mu > 0.0035$~eV/c at the $95\%$ confidence level \cite{Kapner:2006}. These efforts are at
the threshold of the scale inferred from $\Lambda$.

Long-distance gravitational phenomena are also sensitive to such modifications and provide a
tighter bound on $\mu$, the scale of new physics. The key is the limited range of graviton
momenta mediating the gravitational force exerted by a massive body on a test particle.
Considering the deflection of light as an elastic, quantum mechanical scattering process, the
photon energy is conserved but its momentum is redirected. A maximum graviton momentum implies a
maximum deflection angle, and so $|\vec k_{\gamma i} - \vec k_{\gamma f}| \approx 2 k_\gamma \theta
< \mu$, where $k_\gamma$ is the photon momentum. 

We perform a calculation of tree-level photon scattering in linearized quantum gravity. 
We treat the lens as one massive particle, as many constituent particles, or
as the source of an external gravitational field. All approaches yield the same result.
The external field offers the clearest view. The cross section is 
\begin{equation}
\sigma = (2 \pi)^2 \int d^3k_{\gamma f}
\delta(k_{\gamma i} - k_{\gamma f}) 
| \langle k_{\gamma f} |\mathcal{M} | k_{\gamma i}\rangle |^2,
\end{equation}
for a given photon polarization. The Maxwell tensor $T^{\mu \nu}=F^{\mu
\rho}F^{\nu}_{\,\,\rho}-\frac{1}{4}\eta^{\mu \nu}F_{\alpha \beta}F^{\alpha \beta}$ is used in
(\ref{Lint}) to determine the scattering vertex, and the matrix element is calculated in the
external-field approximation, using $h^{\mu\nu}$ for a weak gravitational field due to a point
source of mass $M$. Following Refs.~\cite{Bocca67,DeLogi:1977dp} we obtain 
\begin{eqnarray}
\langle k_{\gamma f} |\mathcal{M} | k_{\gamma i}\rangle &=&
\frac{8 \pi G M}{2 (2 \pi)^2} \sqrt{k_{\gamma f} k_{\gamma i}}
\frac{{\rm e}^{-|\vec{k}_{\gamma f}-\vec{k}_{\gamma i}|/\mu}
}{|\vec{k}_{\gamma f}-\vec{k}_{\gamma i}|^2} \, \Pi(e,k) \quad \cr
\Pi(e,k) &=& \frac{1}{\sqrt{2}}[(\hat e_i\cdot \hat e_f^{*})
(3 - \hat k_{\gamma i}\cdot \hat k_{\gamma i}) \cr
&&\qquad \qquad + (\hat e_f^{*}\cdot\hat k_{\gamma i}) (\hat e_i\cdot\hat k_{\gamma f})]
\end{eqnarray}
where $\hat e$ is the photon polarization vector. Averaging over incoming photon polarizations
and summing over outgoing polarizations, we obtain the differential cross section in the small
angle limit
\begin{equation}
\frac{d\sigma}{d\Omega} =  
\frac{(4 G M)^2 }{(c \theta)^4 } \times {\rm e}^{-2 \theta k_\gamma/\mu} \, .
\label{crosssection}
\end{equation}
In the absence of the cutoff, the cross section has the familiar $\theta^{-4}$ dependence found
in Coulomb scattering. With the cutoff, we interpret the
result to indicate that high-energy photons find a weaker gravitational lens, than low-energy
photons. This stands in contrast with the achromatic nature of lensing in general relativity. 

\begin{figure}[t]
\includegraphics[width=3.25in]{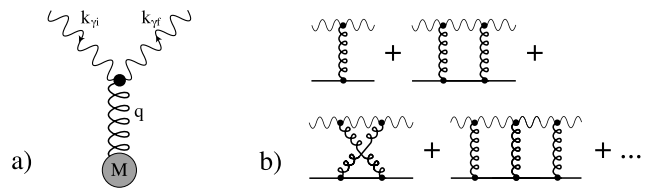}
\caption{a) The Feynman diagram for
the gravitational deflection of light.  
b) The leading ladder and crossed-ladder Feynman diagrams for
graviton exchange are shown.}
\label{figure}
\end{figure}

It is not surprising that gravitational lensing can be described by a tree-level diagram. As with
Coulomb scattering, a tree-level diagram is sufficient to reproduce the classical result. We may
also calculate the contribution of higher-order Feynman diagrams in the {\it eikonal} limit,
wherein the total energy of the colliding particles vastly exceeds the momentum transfer. This
clearly applies to astrophysical gravitational lensing. In perturbative quantum gravity, graviton loop diagrams are responsible 
for the nonrenormalizability of the theory and
lead to a loss of predictive power at high energies. In the eikonal limit, these diagrams are
negligible compared to the series of ladder and crossed-ladder diagrams illustrated in
Fig.~\ref{figure}. As shown in Refs.~\cite{thooft,kabatortiz}, the amplitude for gravitational
scattering of two massive scalar particles can then be summed
to all orders in perturbation theory. In the absence of a cutoff on graviton
momenta, this procedure yields the amplitude multiplied by a divergent phase factor. Since the
cross section depends on $|\mathcal{M}|^{2}$, the Born approximation for the cross-section is
exact. We generalize this result to the case with the cutoff.
We work in the rest frame of the massive scatterer and include an
exponential factor for the momentum cutoff on each graviton propagator. The photon is adequately
treated as a massless scalar in the limit of small deflections. Then,
following Ref.~\cite{kabatortiz}, the scattering amplitude due to an infinite sum of ladder
graphs in the eikonal limit is
\begin{eqnarray}
\label{cdgnonpert1}
\nonumber
i \mathcal{M}=\frac{8\pi M E_{\gamma}}{q^{2}}\,{\rm e}^{-q/\mu}\,
\int_{0}^{\infty} dz~z~J_{0}(z)\times& \\
\left[\left(k_{\rm IR}/\mu+\sqrt{\left(k_{\rm IR}/\mu\right)^{2}+\left(z k_{\rm IR}/q\right)^{2}}
\right)^{4 i \eta}-1\right].
\end{eqnarray}
As in QED, the infrared regulator $k_{\rm IR}$ is necessary because the asymptotic states assumed
were plane waves, rather than Coulombic wave functions. To proceed, we make a series expansion in
small $k_{\rm IR}/\mu$. Then, because $\eta \equiv G M E_{\gamma}\gg 1$, the integral is found to
be well-approximated by
\begin{equation}
i \mathcal{M} = i \mathcal{M}_{{\rm Born,GR}}\,{\rm e}^{-q/\mu}\,
\left(\frac{4 k_{\rm IR}^{2}}{q^{2}}\right)^{2 i \eta}  
\frac{\Gamma(1 + 2 i \eta)}{\Gamma(1 - 2 i \eta)}{\rm e}^{iq/\mu},
\end{equation}
where $\mathcal{M}_{{\rm Born,GR}} = 32 \pi G M^2 E_\gamma^2/q^2$ for the gravitational
scattering of these two scalar particles. This nonperturbative result consists of the
exponentially suppressed Born amplitude with an additional phase which does not affect
the scattering cross-section. Thus our tree-level result is exact in the eikonal limit.

As opposed to multiple graviton exchange in a single scattering interaction, we may also consider
multiple encounters along the particle trajectory. For photons impinging on a
target with an impact parameter $b$, the gravitational interaction time is $\sim b/c$. In
comparison, the interval during which the photon is in the near, scattering zone of the
gravitational lens is $\Delta t \sim b/c$. From the similarity of these time scales, we expect
that the photon will experience but a single scattering interaction. For a non-relativistic
particle of velocity $v$, we expect the interval $\Delta t\sim b/v$ will be much greater than
$b/c$. We thus expect the deflection to be determined by many, successive
single-graviton exchange interactions with the central mass.  Hence, bound systems as well as the
scattering of massive objects, such as satellites or stars, are insensitive to the cutoff since
they exchange lower momentum gravitons at each vertex.

We can also consider the photon deflection as arising from multiple scatting events off the constituent
particles in the deflector mass. In QED, when an electron scatters off a heavy nucleus, it has a
single photon vertex, but each charged nucleon couples coherently to a virtual photon. The
total scattering matrix element is the sum of the matrix elements due to the individual
scatterers \cite{Lax:1951}. If $\mathcal{M}_j$ is the matrix element for the jth scatterer, the
total amplitude is $|\mathcal{M}_{\rm tot}|^2 = \sum_{j} |\mathcal{M}_j|^2 + \sum_{j\neq j'}
\mathcal{M}_j^{*} \mathcal{M}_{j'}$. For $Z$ constituent particles there are $Z$ diagonal terms
and $Z(Z-1)$ off-diagonal terms. Evaluation of the off-diagonal terms requires the correlations
between $j,\,j'$ pairs of particles.  The incoming electron scatters coherently, as is the case
for weak deflections in which the internal momenta of the nucleons are negligible, so the $j$
particles all move with the nucleus zero mode, and the correlations are effectively
delta functions. Upon integration over the phase space to obtain the differential cross section,
the $Z^{2}$ diagonal and off-diagonal terms contribute equally, and so the multiple scattering
approach yields the same result as scattering off the collective nucleus.  

In the case of gravitational deflection, we may consider the deflector mass $M$ as consisting of
$Z$ smaller objects of mass $M/Z$, which includes the gravitational binding energy. For typical
gravitational lens systems, the impact parameter is much greater than the deBroglie wavelength
corresponding to the total momentum transfer. Thus, we are in the limit of coherent scattering,
and as in QED, the same result is obtained whether we employ the point particle or
multiple scattering description. Since the scattered particle has only one vertex, the cutoff
leads to the same constraint on the change in photon momentum, resulting in
Eq.~(\ref{crosssection}) for the cross section. 

To interpret the cross section in terms of a deflection angle, we consider an incident beam of
light at impact parameter $b$. The beam is deflected into an area $d\sigma=~b~db~d\phi$, which
gives us a differential relating $\theta$ and $b$. For small angles, this differential can be
integrated to yield $4 G M/(b c^2) = \theta / F(2\theta k_\gamma /\mu)$ where $ F(x) = 
\sqrt{(1-x){\rm e}^{-x} - x^2{\rm Ei}(-x)}$ and $Ei(x)\equiv -\int_{-x}^\infty {\rm e}^{-t} dt/t$
is the exponential-integral function. Defining $\theta_{GR} \equiv 4 G M/(b c^2)$ for the
standard result without the cutoff, then $\theta/\theta_{GR} = F(2\theta k_\gamma /\mu)$. We note
that the static, frequency-independent metric potential is insufficient to describe the photon's
path past the lensing source when $\theta_{GR} \gtrsim \mu/2 k_\gamma$. It would be necessary to
introduce an effective force into the geodesic equation, based on the modified graviton
propagator. We thus find that the deflection is half the standard prediction 
when  $2 \theta k_\gamma /\mu \sim 1$. In the limit $\theta \ll \mu/2k_\gamma$, $F\to 1$, but for
$\theta \gtrsim \mu/2 k_\gamma$ the deflection angle is suppressed. Hence, \textit{we would
expect a dearth of gravitationally lensed images of high-frequency light if there were a cutoff
in graviton momentum.}

Numerous gravitational lens systems have been observed from radio to x-ray frequencies. The
tightest constraint to $\mu$ comes from x-ray observations of the gravitationally lensed system
Q0957+561 \cite{Chartas:2002}. For this lens system, image A due to the quasar at $z=1.4$ appears
$5.2''$ away from the primary lensing galaxy at $z=0.36$ \cite{castles}. Using the
angular-diameter distances to the source and from lens to the source, $D_S,\,D_{LS}$, to
reconstruct the lensing geometry, we estimate a deflection angle of $\theta = 5.2" \times
D_S/D_{LS} = 7.8''$. The lens image locations are unchanged for $E_\gamma < 5$~keV
\cite{Chartas:2005}, which yields the lower bound $\mu > 0.38$~eV/c.  This result pushes the
threshold for departures from the Newtonian force law down to $0.5~\mu$m.

This lower limit is nearly two orders of magnitude higher than, and therefore rules out, the
cutoff inspired by the cosmological constant with $N>1$. If $N \ll 1$ perhaps due to a
cancellation of bosonic and fermionic contributions, then agreement is still possible. We have
also tried other forms for the cutoff, including a Gaussian and a sharp power law
and find that our results do not change appreciably. This bound may also constrain dark energy
models, where such a cutoff prevents the spontaneous decay of the vacuum into phantom or ghost
particles \cite{Carroll:2003st,Cline:2003gs,Hsu:2004vr,Kaplan:2005rr}.  We caution the reader
that our results only apply to effective theories in which gravity weakens above the cutoff
scale in a way described by the implementation of the cutoff function $f(q)$.  A tighter
constraint may be obtained in the future from hard x-ray or gamma-ray observations of lens
images. 

It is instructive to compare our graviton momentum cutoff with a similar cutoff in the
electron-phonon interaction. In metals, the phonon plays an important role in the dynamics of
conduction electrons, conveying an attractive long-range interaction between electrons, which
partially cancels the Coulomb interaction. The phonon has an effective width or frequency which
characterizes the response time of the ion lattice, above which the phonon interaction is
suppressed.  The bare pseudo-potential extracted from the electron-phonon matrix element must be
dressed by frequency-dependent factors which include the limited phonon-response, in order to
produce an accurate picture of the electron dynamics (e.g. \cite{Scalapino:1969}). By analogy
with the phonon,  we expect the effective width of the graviton to lead to a dramatic change in
the behavior of gravitational scattering, shifting the boundary between classical and quantum
gravitational interactions. Tree-level amplitudes, which are usually regarded as classical due to
the absence of any $\hbar$ factor, are quantum-corrected by the presence of the phenomenological
scale $\mu$. We expect that the static gravitational potential will be of limited use, since it
may not fully capture the effects of the limited graviton response on kinematics. 

We note that a graviton cutoff would lead to a suppression of the spectrum of inflationary
gravitational waves. The highest frequency graviton modes allowable by the cutoff enter the
horizon when $H \sim c \mu/\hbar$, at which time the cosmic temperature is $\sim 2$~TeV for a
cutoff based on the magnitude of $\Lambda$. These waves redshift down to a frequency $\sim 2
\times 10^{-4}$~Hz by the present day. Hence, there would be no inflationary gravitational waves
in the frequency range of the proposed Big Bang Observer \cite{BBO:2004} satellite gravitational
wave detector. 

We have explored the consequences of a simplistic treatment of the cosmological constant problem.
Here, with the introduction of the momentum scale $\mu$, the classical regime is restricted  to
soft interactions with low momentum transfers; hard scattering must take into account the 
suppression factor on the graviton propagator. One may expect a cutoff to play some role in
separating the high energy and low energy domains of the underlying, fundamental theory of
gravity.  At energy scales above the cutoff, gravity may weaken and then lensing imposes an
important bound. 

\begin{acknowledgments}
We thank George Chartas for sharing his Chandra results. We thank the TAPIR group
and Mark Wise at Caltech for useful discussions. R.C. was supported in part by NSF AST-0349213
and Research Corporation RI-0887 at Dartmouth. 
\end{acknowledgments}
 
%%%%%%%%%%%%%%%%%%%%%%%%%%%%%%%%%%%%%%%%%%%%%%%%%%%%%%%%%%%%%%%%%%%%%%%%%%%%%%%%%%%%%%%%%%%%%%%%%%

\end{document}